\documentclass[preprint]{emulateapj}

\newcommand{\HI}{H \textsc{i} }
\newcommand{\HII}{H \textsc{ii} }

 \usepackage{apjfonts}

\begin{document}

\title{The Abundance Gradient in the Extremely Faint Outer Disk of NGC 300}

\shorttitle{The Abundance Gradient in the Disk of NGC 300}
\shortauthors{Vlaji\'c, Bland-Hawthorn \& Freeman}

\author{M. Vlaji\'c} 
\affil{Astrophysics, Department of Physics, Keble
Road, University of Oxford, Oxford OX1 3RH, UK}
\email{vlajic@astro.ox.ac.uk}

\author{J. Bland-Hawthorn} 
\affil{Institute of Astronomy, School of
Physics, University of Sydney, NSW 2006, Australia}

\and

\author{K. C. Freeman}
\affil{Mount Stromlo Observatory, Private Bag, Woden, ACT 2611, Australia}

\begin{abstract}
In earlier work, we showed for the first time that the resolved
stellar disk of NGC~300 is very extended with no evidence for
truncation, a phenomenon that has since been observed in other disk
galaxies. We revisit the outer disk of NGC~300 in order to determine
the metallicity of the faint stellar population. With the GMOS camera
at Gemini South, we reach $50\%$ completeness at ($g'$, $i'$) =
($26.8-27.4$,$26.1-27.0$) in photometric conditions and 0.7"
seeing. At these faint depths, careful consideration must be given to
the background galaxy population. The mean colors of the outer disk
stars fall within the spread of colors for the background galaxies,
but the stellar density dominates the background galaxies by
$\sim2:1$.  The predominantly old stellar population in the outer disk
exhibits a negative abundance gradient -- as expected from models of
galaxy evolution -- out to about $10$ kpc where the abundance trend
changes sign. We present two scenarios to explain the flattening, or
upturn, in the metallicity gradient of NGC~300 and discuss the
implication this has for the broader picture of galaxy formation.
\end{abstract}


\section{INTRODUCTION}
\label{sec:introduction}

Simulations of galaxy formation in cosmological context reveal that
the process of galaxy assembly is expected to leave an imprint on the
characteristics of stars in the faint outer regions of disk galaxies
\citep{bullockjohnston05}.  Due to their long dynamical timescales
these regions retain the fossil record from the epoch of galaxy
formation in the form of spatial and kinematic distributions, ages and
chemical abundances of their stars \citep{freemanbh02}.  These are
also the regions where the effects of stellar radial mixing are
expected to be most prominent \citep{roskar08,schoenrich08}.  The
structure of the outer parts of galactic disks is therefore central to
our understanding of the formation and evolution of disk galaxies.

The studies of galactic disks have identified three basic classes of
surface brightness profiles in galaxies \citep{pohlen07}. Truncations
in surface brightness profiles, characterized by the smooth break
between the inner shallower and the outer steeper exponential, have
been known for more than two decades and detected in a large number of
galaxies \citep{vanderkruit79,degrijs01,pohlen02}. Despite this, the
origin of the break remains unclear. The currently favored model for
the explanation of these ``sub-exponential'' truncations is the star
formation threshold scenario (\citealp{kennicutt89,schaye04,roskar08},
but see \citealp{vanderkruit07}). However, the threshold
gas densities predicted in these models are low \citep[$3-10$
  $M_{\odot}$ pc$^{-2}$;][]{schaye04} compared to the observed surface
brightnesses at the radius of the break which correspond to stellar
mass densities of $20-2000$ $M_{\odot}$ pc$^{-2}$ \citep{erwin07}. The
second class of surface brightness profiles, so called ``upbending'' or
``super-exponential'' breaks, have been explained by a scenario
involving minor mergers, supported both observationally
\citep{ibata05} and by N-body/SPH simulations \citep{younger07}. Until
recently, it was believed that all galaxy disks undergo truncation
near the Holmberg radius (26.5 mag arcsec$^{-2}$) and that this was
telling us something important about the collapsing protocloud of gas
that formed the early disk \citep{vanderkruit79}. This picture was
challenged by our earlier finding \citep{blandhawthorn05} of a classic
exponential disk \citep{freeman70} with no break in NGC~300, out to
the distances corresponding to $10$ disk scale lengths in this
low-luminosity late type spiral. 

The existence of stars at such low stellar surface densities
\citep[$0.01$ $M_{\odot}$ pc$^{-2}$;][]{blandhawthorn05} in the outer
spiral disk has provided the motivation for this work. We seek to
learn which physical process are responsible for the formation of the
outer disks, and what the nature of the stellar populations in this
rarefied medium is. To this end, we obtained additional multi-band
photometry ($g'$,$i'$) with Gemini/GMOS which allowed us to explore
ages and metallicities of the stars in the outer regions of NGC~300. 

A number of studies have addressed the question of star formation in
outer galactic disks.  \citet{davidge03,davidge06,davidge07} find
young and intermediate-age stars at large galactocentric distances in
NGC~2403, NGC~247 and M33. \citet{barker07} discovered young (less
than $500$ Myr old) main sequences (MS) stars in the outer regions of
M33, beyond the break radius at $4.5$ scale lengths
\citep{ferguson07}. The existence of stars less than $0.5$ Gyr old
would suggest a very recent star formation episode in the outer disk
of M33 as there probably has not been sufficient time for these stars
to be scattered from the inner disk and migrate outward. The same
study also finds that the scale lengths of stellar populations in M33
increase with age. This is in contrast to the common scenario of
inside-out galaxy formation in which the scale length of stars
increase as the disks builds up to its present size, and it is
expected that the mean age of the stellar disk decreases with radius
\citep{larson76,matteuccifrancois89,chiappini97,naabostriker06,munozmateos07}.
These findings emphasize how little is understood about the formation
of outer disks of spirals and call for deep photometric data on more
pure disk galaxies.

The Sculptor Group of galaxies, of which NGC~300 is a member, is one
of the nearest and largest ensembles of disk galaxies outside the
Local Group. Its galaxies are believed to be more or less isolated and
as such are an ideal sample for our proposed study. NGC~300 is an
almost pure disk galaxy (bulge light $<2\%$), with mild inclination
($i=42^\circ$), which makes it well suited for stellar content
studies. It lies at high galactic latitudes and therefore has low
foreground reddening of $E(B-V)=0.011-0.014$ mag \citep{schlegel98},
which translates into $E(g'-i')=0.018-0.024$ mag. The high galactic
latitude of Sculptor galaxies is also important because the
contamination from the foreground Galactic stars is minimal at these
high latitudes. In many respects NGC~300 is similar to the local
spiral M33 \citep{blairlong97}. The optical disk scale length has been
measured as $2.23$ kpc in B$_J$ \citep{carignan85} and $1.47$ kpc in I
\citep{kim04}. \citet{tikhonov05} and \citet{butler04} find that the
stellar population is predominantly old and metal-poor, with no
evidence for significant change in metallicity with time. The study of
\citet{puchecarignan90} finds in NGC~300 an extended \HI disk, which
is severely warped in the outer parts ($10'<r<20'$). The warp in \HI
disk has often been linked to the truncations in the stellar disk
\citep{vanderkruit07}.

We adopt a distance modulus of 26.51 for NGC~300. This is a
weighted-mean of distances determined from the Cepheid Variables
\citep{freedman01,gieren04} and the tip of the red giant branch
\citep{butler04,tikhonov05}.

The layout of the paper is as follows. We summarize the details of the
observations and the data reduction in \S\ref{sec:observations} and
describe the photometric measurements and artificial stars test in
\S\ref{sec:photometryandcompleteness}. The color-magnitude diagrams,
star counts and surface brightness profiles, and the metallicity
distribution function of NGC~300 are presented in
\S\ref{sec:results}. Finally, we discuss our results in 
\S\ref{sec:discussion}.  

\section{OBSERVATIONS AND DATA REDUCTION}
\label{sec:observations}

\begin{figure}[]
\plotone{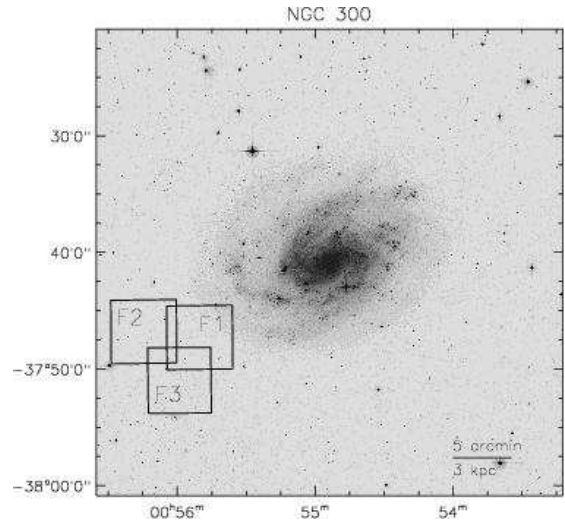}
\caption{The Digitized Sky Survey (DSS) wide-field
  image of NGC~300. (North is to the top, east is to the left.)
  Squares mark the positions of GMOS $5'\!.5\times5'\!.5$ field of
  view.\label{fig1}}
\end{figure}

The deep $g'$ and $i'$ images of three fields in the outskirts of
NGC~300 were obtained using the Gemini Multi Object Spectrograph
(GMOS) on Gemini South telescope as a part of the program
GS-2005B-Q-4; in total $34$ hours of data have been taken.  
The observations are summarized in Table~\ref{obs}. The NGC~300 fields
span a major axis distances of $7-16$ kpc and were made to overlap
slightly to allow for the check of the relative photometric
calibration between the fields. The inner-most field was placed at the
optical edge identified in the Digitized Sky Survey (DSS) to provide
continuity with earlier work. The locations of the fields are marked
in the Figure~\ref{fig1}. The field of view in a single GMOS image
is $5'\!.5$ on a side; at the distance of NGC~300 $1'$ corresponds to
$0.6$ kpc. 

\begin{deluxetable*}{ccccccc}
\tabletypesize{\scriptsize}
\tablecaption{GMOS Observing Log.\label{obs}}
\tablewidth{0pt}
\tablehead{
\colhead{Target} & \colhead{RA (J2000)} & \colhead{DEC (J2000)} & \colhead{Date} & \colhead{Filter} & \colhead{Exposure (s)} & \colhead{FWHM ($''$)}}
\startdata
NGC300 1 & 0 55 50 & -37 47 18 & 2005 Sep 7-8 & $g'$ & 13$\times$600 &
0.71 \\ 
... & 0 55 50 & -37 47 24 & 2005 Aug 4 & $i'$ & 22$\times$600 & 0.66 \\ 
NGC300 2 & 0 56 15 & -37 46 45 & 2005 Aug 8-10 & $g'$ & 13$\times$600
& 0.72 \\
... & 0 56 15 & -37 46 45 & 2005 Aug 8-10 & $i'$ & 22$\times$600 & 0.67 \\ 
NGC300 3 & 0 55 59 & -37 50 59 & 2005 Oct 10-11, 31 & $g'$ &
13$\times$600 & 0.82 \\
... & 0 55 59 & -37 50 53 & 2005 Sep 8, Oct 8-10 & $i'$ & 22$\times$600 &
0.72 \\
\enddata
\tablecomments{Units of right ascension are hours, minutes, and
  seconds, and units of declination are degrees, arcminutes, and
  arcseconds.} 
\end{deluxetable*}

The GMOS detector is a mosaic of three $2048\times4608$ EEV CCDs
\citep{hook04}. The raw pixels were binned $2\times2$ during readout to
provide a better match of the image sampling to the seeing, producing
the pixels of $0''\!\!.146$ in size. The average image quality of the
data is $0''\!.7$ FWHM in $g'$ and $0''\!.6$ FWHM in $i'$.

For each field, $13\times600$ s exposures were taken in $g'$ and
$22\times600$ s exposures in $i'$ band. Our goal was to achieve
$3\sigma$ photometry in $g'$ and $i'$ for a G giant star with
(V$=27.5$, B$=28.5$, I$=26.5$). The data were obtained in a dithered
pattern, which allowed for the gaps between the individual GMOS CCDs
to be filled when constructing the final image.

The data were reduced using the standard IRAF/Gemini routines. Due to
the known problems with GMOS bias frames, we inspected all bias frames
and discarded those in which the counts varied significantly between
the subsequent frames. For the same reason, we decided to have the
overscan regions subtracted in the reduction process. The processing
pipeline included creating the master bias and flat field frames
(\texttt{gbias}, \texttt{giflat}), bias subtraction and flat fielding
(\texttt{gireduce}), mosaicking of individual GMOS CCDs into a single
reference frame (\texttt{gmosaic}), and finally combining the dithered
exposures into a final image (\texttt{imcoadd}). In addition, $i'$
data were affected by fringing and the fringes needed to be
removed. The master fringe frame was created for each field from the
individual reduced image frames (\texttt{gifringe}) and subtracted
from the individual images before coadding (\texttt{girmfringe}).

\begin{figure*}[]
\plotone{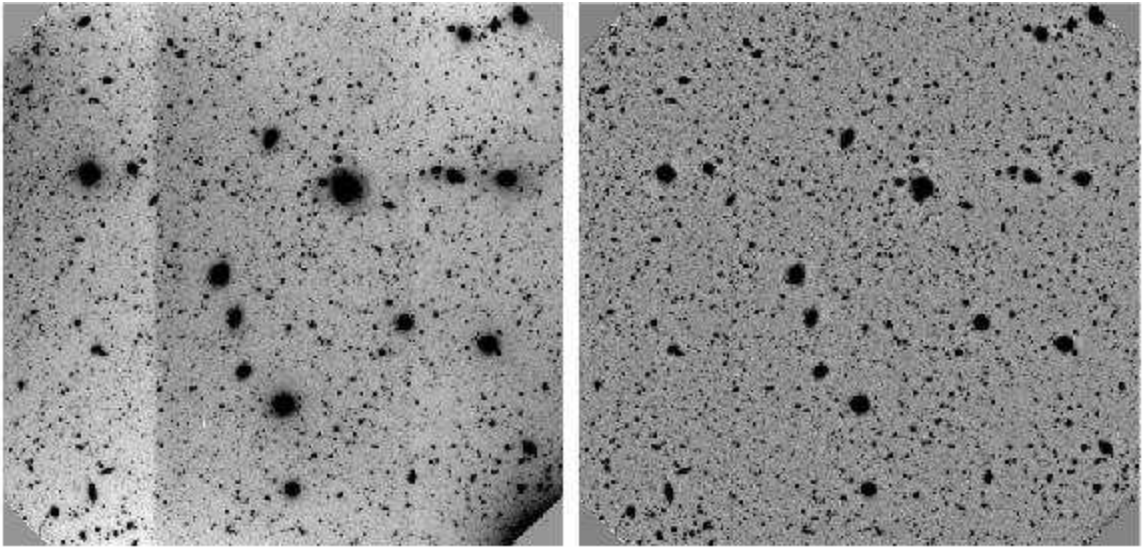}
\caption{The reduced $g'$ images of the NGC~300 F2 field. Left
  panel: the image after the standard reduction procedure has been
  applied (see text for details). Right panel: the image after
  applying the additional fringe removal procedure. This additional 
  reduction step significantly improved the consistency of the
  background between the three GMOS CCD chips.\label{fig2}}
\end{figure*}

After the reduction procedure, the artifacts of mosaicking the GMOS
CCDs into a single image were still present in $g'$ data, visible as a
background offset between the chips in the final images. The level of
the offset was of the order of $\sim1\%$ of the background, but was
clearly evident upon visual inspection, as shown in the left panel of
Figure~\ref{fig2}. In order to attempt to remove these effects, we
applied the same fringe removal procedure on our $g'$ data as
described above for $i'$ images. The procedure improved the background
consistency between the chips; the example of the result for one of
the fields is shown in the right panel of Figure~\ref{fig2}.

\section{PHOTOMETRY AND COMPLETENESS ANALYSIS}
\label{sec:photometryandcompleteness}

\subsection{Photometry}
\label{sec:photometry}

The final $g'$ and $i'$ band GMOS images were analyzed using the
DAOPHOT/ALLSTAR software suite \citep{stetson87}. For each field
DAOPHOT produced object catalogues and performed aperture photometry
in a series of circular apertures. About $100-200$ relatively bright
and isolated stars in each field were chosen as PSF stars and used to
iteratively compute the point spread function (PSF) for the field.
All PSF stars candidates were examined visually within the
IRAF/daophot package and their radial, contour and mesh profiles were
inspected for any indications of their non-stellar nature.  After the
PSF stars were subtracted within the standalone DAOPHOT suite, the
positions of subtracted PSF stars were inspected again using the above
indicators.  All PSF star which were not subtracted cleanly were
excluded from the subsequent PSF calculations.  In addition, stars
which had PSF subtraction errors more than $3\sigma$ away from the
mean value, as reported by standalone DAOPHOT package, were also
excluded.  The next iteration PSF was calculated from the image in
which, within the fitting radius of each PSF star, all but PSF stars
were subtracted. The procedure described above was repeated once more
to derive the PSF; finally, ALLSTAR was used to fit this PSF to the
objects in the catalogues and determine their PSF photometry.

In addition to the information on the object magnitude and the related
error, ALLSTAR provides information on the quality of PSF fitting for
each star through the parameters \textsc{chi} and
\textsc{sharp}. \textsc{chi} represents the ratio of the observed
pixel-to-pixel scatter in the fitting residuals to the expected
scatter. It is expected for the values of \textsc{chi} to scatter
around unity when plotted against the derived magnitude
\citep[cf. Figure 28 of][]{stetsonharris88}. In our analysis however,
the values of \textsc{chi} are scattered around $\sim0.4$. This may be 
indicative of the incorrect values for the gain and readout noise used
in the analysis (P. Stetson, private communication). The
correspondence with Gemini staff confirmed that the gain and readout
noise values used are indeed correct and the reason for the invalid
\textsc{chi} values in ALLSTAR remained unclear. However, any
photometric errors introduced by potentially incorrect gain and
readout noise values will be taken into account through our artificial
star tests which we use to estimate the observational uncertainties in
the data.

The aperture corrections were calculated using the DAOGROW software
\citep{stetson90}, which performs a sophisticated growth-curve analysis
to derive a "total" magnitude for each star. For each field,
multiple-aperture photometry was done on the image frames from which
all but PSF stars have been subtracted. The weighted mean difference
between the PSF-based magnitude (from ALLSTAR) and the ``total''
magnitude (from DAOGROW) of the PSF stars was adopted as ``aperture
correction'' and this correction was applied to PSF-based magnitudes
of all stars in the frame. The uncertainties in the aperture
correction over all bands and fields were $0.006-0.008$ (the quoted
uncertainty is the standard error of the mean). 

The photometric calibration for the GMOS data was established using
the \citet{landolt92} standard star fields.  The $g'$ and $i'$
magnitudes of standard stars were calculated using the relations from
\citet{fukugita96}. In total, $6$ Landolt fields were observed
multiple times over the two months run, with the total of $\sim140$
standard stars observations. The total observed magnitudes of standard
stars were measured using DAOGROW and the zero point was calculated as
the difference between the observed magnitude corrected for the
exposure time and the standard magnitude. Zero points varied by $0.02$
between the standards stars observations on the different nights of
the same month, or by $0.12$ and $0.07$ in $g'$ and $i'$,
respectively, over the two month period.  Photometry was transformed
onto a single calibration by comparing the magnitudes of stars in the
overlapping regions, taking as a reference the field with the best
calibration of the three (F1 in i' and F2 in g', both observed during
the August 2005 run).  Results of the completeness analysis described
below were used to confirm that the scatter of the photometry in the
overlapping regions agrees with the scatter predicted from the
artificial star tests.

\subsection{Completeness Analysis}
\label{sec:completeness}

When interpreting stellar photometry of faint stars in crowded fields,
it is important to understand the observational uncertainties inherent
in the data.  It is a standard procedure to use artificial star tests
to estimate the completeness of the data and assess the effects the
crowding has on the calculated magnitudes and their errors. In the
artificial star test a set of stars with known magnitudes is added to
the frame, and the frame is analyzed using the standard data reduction
pipeline to recover the photometry of all stars, real and artificial.

\begin{figure}[]
\plotone{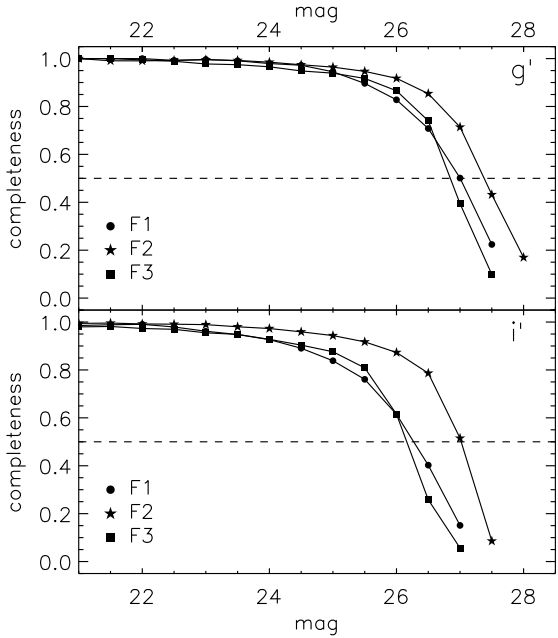}
\caption{Completeness curves for the observed NGC~300
  fields, in $g'$ (top panel) and $i'$ (bottom panel). The horizontal
  dashed line marks the $50\%$ completeness limit.\label{fig3}}
\end{figure}

To be able to estimate the data completeness robustly, a large number
of stars has to be added to the images and analyzed. This number
should be significantly larger than the number of real stars detected
in the image. At the same time, it is important not to affect the
crowding of the images significantly, as that will result in the
unrealistic estimates of completeness. As a compromise between the
crowding in the field (where a small number of artificial stars is
optimal) and the computing efficiency (where a large number of
artificial stars is favorable as it decreases the total number of
images that need to be analyzed), the number of stars that we add to
each frame is $5-10\%$ of the number of detected stars in the
frame. We then run a hundred tests on each image to obtain
statistically robust estimate of the completeness and observational
uncertainties.

In order to sample the crowding conditions in the frame correctly,
artificial stars injected into the image must not overlap. To make
certain that there is no overlap between the artificial stars, we add
them to the frames in a regular grid. The grid has 30 stars on a side
(i.e. 900 in total), with a separation of $69$ pixels ($10''$) between
neighboring stars. Our choice of the grid and cell size is guided by
the considerations explained above and the need for the whole extent
of the image to be sampled in each run.  A total number of stars added
to each frame over $100$ artifical star test runs is $\sim10000$ per
frame. The grid origin has been randomly offset between the runs in
order to enable sampling of the whole extent of the image. The same
grid setup is used to add artificial stars to one ($g'$,$i'$) pair of
images.

To assess the color effects it is important that the colors assigned
to artificial stars correspond to the realistic stellar
populations. To ensure that the magnitude and color distribution of
our artificial stars matches magnitudes and colors in the original
data frames, we construct artificial photometry by randomly selecting
stars from the original CMDs and assigning their $g'$ and $i'$
magnitudes to artificial stars.

Artificial stars with the positions and magnitudes determined as
described above are added to the frames using the DAOPHOT/ADDSTAR
routine. The resulting images are analyzed with the same data 
reduction pipeline as the original science frames, using the PSF
calculated for each field as described above. A star is considered
recovered if it is detected in both $g'$ and $i'$ frame and if the
difference between the injected and recovered magnitude in both bands
is $<0.5$ mag. Completeness as a function of magnitude is calculated
as the ratio of the number of recovered and injected stars in a given
$0.5$ magnitude bin. Completeness curves for three NGC~300 fields are
shown in Figure~\ref{fig3}. We compute the completeness in
terms of the instrumental magnitude and then transform them using the
appropriate photometric calibration equations into the completeness in
terms of the standard magnitude. Our photometry is $50\%$ complete
down to ($g'$,$i'$)=($26.8-27.4$,$26.1-27.0$). 

\begin{figure}[]
\plotone{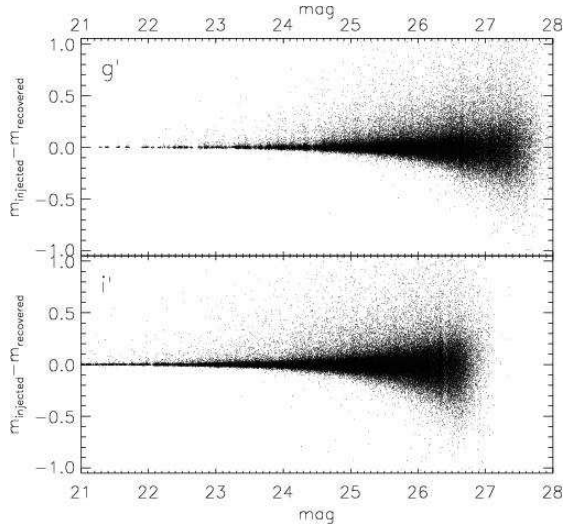}
\caption{The difference between the injected and recovered magnitudes
  of all recovered artificial stars, in $g'$ (top panel) and $i'$
  (bottom panel), for one of the observed NGC~300 fields. Only the
  stars with the difference between the input and output magnitude
  smaller than $0.5$ were considered recovered when calculating
  completeness curves shown in Figure~\ref{fig3}. Other
  NGC~300 fields show similar behavior.\label{fig4}} 
\end{figure}

\begin{figure}[]
\plotone{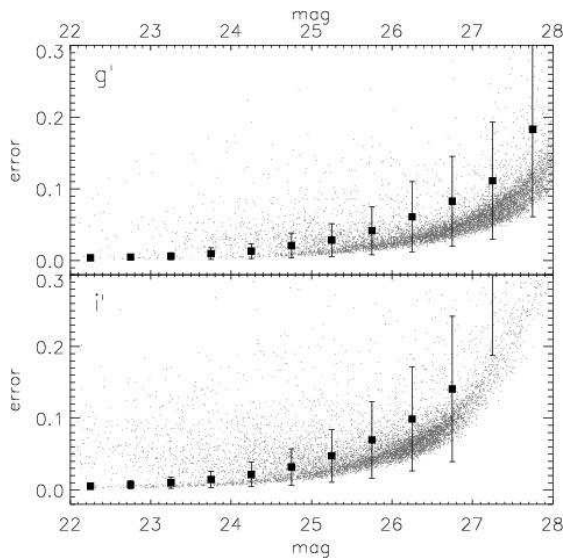}
\caption{Photometric errors as given by ALLSTAR (grey points) and
  calculated from the artificial star tests (black squares); top panel
  shows the results for $g'$ band, bottom panel for $i'$ band
  data. Photometric uncertainties are derived from the artificial star
  tests as the mean difference between the input and output magnitudes
  of artificial stars in $0.5$ magnitude bins.  Error bars give the
  scatter in each bin.\label{fig5}}
\end{figure}

To check the accuracy of the photometry we calculate the difference
between magnitudes of injected and recovered artificial stars.
Figure~\ref{fig4} shows the difference between the input and output
magnitudes as a function of the input magnitude for one of the
observed fields.  Over the most of the magnitude range in $g'$ and
$i'$ the photometric error with which the stars are recovered is
$<0.3$ mag.  There is a systematic trend over the whole magnitude
range, with stars being recovered brighter than their true magnitude.
Assuming that this effect is purely due to blending, we estimate the
blending fractions to be $5-10\%$ in F2 and F3 and $\sim20\%$ in F1.
However, the calculated blending fractions are an overestimate because
at the faint end random fluctuations in the unresolved background may
scatter faint stars below the detection limit, with only the stars
lying on the positive background fluctuations being recovered.

We use artificial star tests to derive uncertainties of our
photometry.  We adopt as photometric errors the mean of the absolute
value of the difference between the input and output magnitudes of
artificial stars. Figure~\ref{fig5} shows the errors reported by
ALLSTAR (points) and photometric uncertainties calculated from the
artificial star tests as described above (squares), as a function of
input magnitude. The error bars are the scatter in each $0.5$ mag
bin. At the faint end the errors derived from the artificial star
tests are $\sim0.04$ mag larger than the errors given by ALLSTAR.

\begin{figure}[]
\plotone{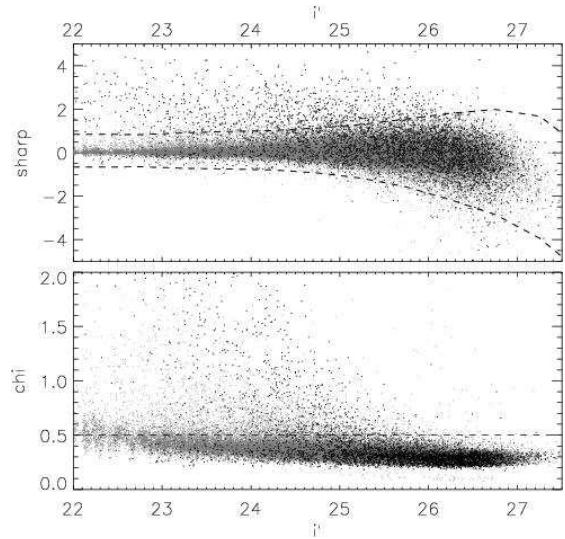}
\caption{Values of the \textsc{chi} (bottom panel) and \textsc{sharp}
  (top panel) parameters as reported by ALLSTAR for all recovered
  artificial stars (grey) and the real stars (black). Dashed lines are
  the cuts in \textsc{chi} and \textsc{sharp} used to discard spurious
  detections.\label{fig6}} 
\end{figure}

In addition to estimating completeness and photometric errors, we use
the artificial stars tests to discard the non-stellar and spurious
detections. Due to the non-standard \textsc{chi} output of the ALLSTAR
routine (see \S\ref{sec:photometry}), we are not able to apply
commonly used cuts in \textsc{chi} parameter to reject false
detections. With the artificial star tests however, we are able to use
the information on \textsc{chi} and \textsc{sharp} of the artificial
stars to set the criteria for removing the false detections from the
data. In Figure~\ref{fig6} we show the distribution of \textsc{chi}
and \textsc{sharp} parameters for the real stars (black points) and
all recovered artificial stars (grey points) as a function of
magnitude, for one of the observed NGC~300 fields.  In order to remove
non-stellar objects from the sample, we employ cutoffs based on the
values of \textsc{sharp} parameter of the recovered artificial stars.
We divide the data into 0.5 magnitude wide bins and discard objects
that are more than $3\sigma$ away from the mean value of
\textsc{sharp} in a given bin.  The dashed curve in the top panel of
the Figure~\ref{fig6} shows the extent of the allowed \textsc{sharp}
values.  In addition, we use a constant \textsc{chi} limit and remove
from the stellar catalogues all stars with \textsc{chi}$>0.5$.

\section{RESULTS}
\label{sec:results}

\subsection{Color-Magnitude Diagram}
\label{sec:cmd}

\begin{figure*}[]
\plotone{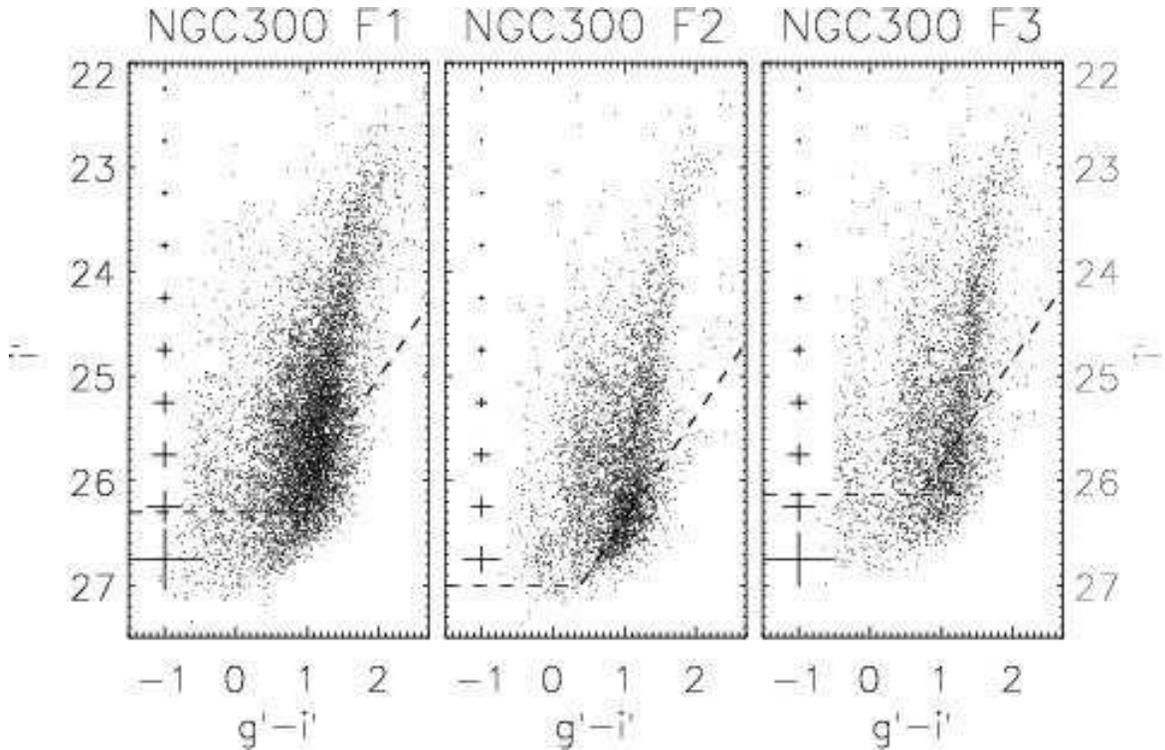}
\caption{Color-magnitude diagrams (CMD) of the three NGC~300
  fields. Error bars show the photometric uncertainties from the 
  artificial star tests at a given $i'$ magnitude and the color of
  $g'-i'=1$. The dashed lines indicate the $50\%$ completeness
  limit.\label{fig7}}
\end{figure*}

We show the ($g'-i'$,$i'$) color-magnitude diagrams (CMDs) of the
three observed NGC~300 fields in Figure~\ref{fig7}. The most prominent
component of these CMDs is the well-populated red giant branch
(RGB). This feature is typical of a system composed of a stellar
population older than $1$ Gyr. In all observed fields the red giants
branch stars make up $\sim50\%$ of all stars detected in the field. A
small number of red stars is found above the tips of the theoretical
red giant branch tracks (Figure~\ref{fig8}); these stars could be
intermediate-age asymptotic giant branch stars and young red
supergiants. Also marked on the CMDs in Figure~\ref{fig7} are the
$50\%$ completeness limit and the photometric errors at a given $i'$
magnitude and $g'-i'$ color of $1$, as determined from the artificial
stars tests (\S\ref{sec:completeness}). 

In addition to the dominant old and red stellar population, we detect a
sparsely populated main sequence with $g'-i'$ colors from $-1$ to
$\sim0$. This feature contains $10\%$ of the stars detected in the
outer disk of NGC~300 and most likely represents main-sequence stars
with the ages of a few hundred Myr.

\subsection{Star Counts Profile}
\label{sec:radialprofile}

In our previous study of the outer regions of NGC~300
\citep{blandhawthorn05} we used $r'$ star counts to trace the faint 
stellar disk out to $15$ kpc ($25'$). Now with the multi-band
photometry in hand we can refine our radial profile to target distinct
stellar populations in the outer disk of NGC~300.

\epsscale{0.8}
\begin{figure}[]
\plotone{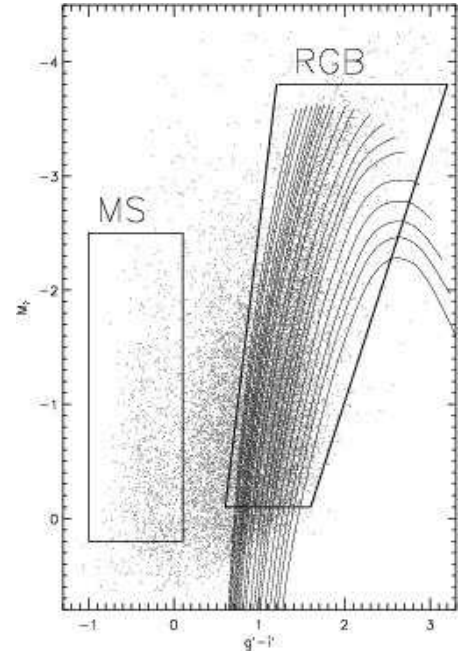}
\caption{Color magnitude diagram of the three NGC~300 fields, with the
  Victoria-Regina \citep{vandenberg06} set of [$\alpha$/Fe]$=0.0$, $8$
  Gyr stellar evolutionary tracks overplotted. The model grid has the
  metallicities in the range [Fe/H]=[$-2.31$,$0.00$]. Two regions 
  marked in the figure were used to select stars belonging to red
  giant branch (RGB) and main sequence (MS) population.  
  \label{fig8}}  
\end{figure}

\begin{figure*}[]
\plotone{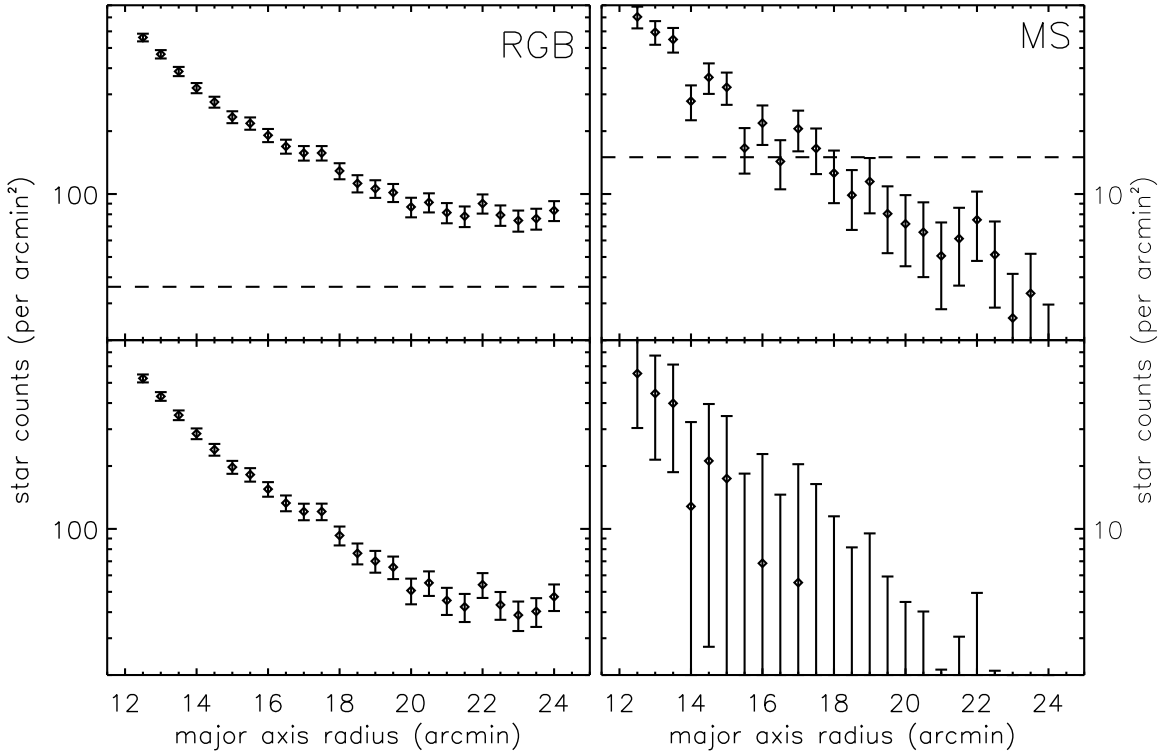}
\caption{Star counts profile of the red giant branch (left panels) and
  main sequence stars (right panels) in the outer disk of
  NGC~300. Stars inside the regions marked in
  Figure~\ref{fig8} were used to calculate the profile.  The
  top panels show the profiles before the contribution of the faint
  unresolved galaxies has been subtracted.  The dashed line
  corresponds to the predicted number of background galaxies (per
  arcmin$^2$) with the colors and magnitudes within our RGB and MS
  selection boxes.  The bottom panels show the star counts profiles
  corrected for the background galaxies contribution.\label{fig9}}
\end{figure*}

We derive the radial profile of the red giant branch and main sequence
stars by counting the stars in elliptic annuli with the ellipticity
(inclination) and position angle corresponding to that of NGC~300
($\mathrm{PA}\!=\!110^\circ$, $i=42^\circ$). In
Figure~\ref{fig8} we reproduce the color-magnitude diagram of
the NGC~300 F1 field and define the selection boxes we use to select
RGB and MS stars. To account for the crowding effects we compute
completeness as a function of galactocentric distance using the output
of the artificial stars tests described in
\S\ref{sec:completeness}. This is calculated as a ratio of the number
of recovered to injected stars in each $0.5'$ radial bin and used to
correct star counts profiles. The resulting completeness-corrected
profiles of RGB and MS stars are shown in the top panels of the
Figure~\ref{fig9}.

\epsscale{1}

In order to be able to derive the intrinsic star counts profile it is
crucial to as reliably as possible estimate the contribution of the
contaminant sources -- dwarf stars in the Galaxy and faint unresolved
background galaxies that get mistaken for NGC~300 stars.  Given the
high galactic latitude of NGC~300 ($b=-79.4^\circ$) we do not expect
the contribution from galactic dwarfs to be significant.  Besancon
models \citep{robin03} predict 26 stars per GMOS field of view in the
magnitude and color range corresponding to our RGB selection box; this
accounts for $0.5-1\%$ of all detected objects in a single field with
the same color-magnitude criteria.

The online \textsc{GalaxyCount} calculator \citep{ellisbh07} is a
useful tool for estimating galaxy counts in deep wide-field images.
It had confirmed our earlier finding that even when the contribution
from the faint unresolved galaxies is accounted for, the exponential
disk of NGC~300 as traced by $r'$-band star counts extends out to 10
disk scale lengths \citep{blandhawthorn05, ellisbh07}.  However, we
now derive star counts profiles for different stellar populations
which populate distinctive regions in the color-magnitude space
(Figure~\ref{fig8}) and require information on the color, as well as
magnitude, of the background galaxy population.  Since
\textsc{GalaxyCount} at present does not include the color
information, we can only estimate the number counts predicted in $g'$
and $i'$ bands separately.  For example, in the case of our RGB
selection box (Figure~\ref{fig8}) \textsc{GalaxyCount} predicts
$82-114$ galaxies in the magnitude range $g'=23.9-28.0$ and $114-140$
galaxies in the range $i'=22.7-26.4$. To estimate the galaxy number
counts we use the completeness curves derived separately for each
field; the range in the number counts estimates reflects the
differences in the parameters describing the completeness curves for
different fields.  However, the number counts predicted by
\textsc{GalaxyCount} are an overestimate of the true contamination
within our RGB selection region because (1) they are derived without
any restrictions on the color of the background galaxy population and
(2) a number of contaminating galaxies has already been somewhat
reduced by the use of the \textsc{sharp} statistic.

For the reasons described above we chose to use the galaxy counts from
the William Herschel Deep Field \citep[WHDF,][]{metcalfe01} to
estimate the contamination from the faint background galaxy
population. The main caveat with this kind of approach is that the
WHDF survey is not as deep as our data and some assumptions need to be
made in order to estimate the number counts of galaxies in our data
using a shallower survey. We calculate the $i'$-band number counts of
the galaxies falling within our color selection criteria and fit
linearly the (log of) differential number counts in $0.5$ mag bins in
the range covering $3$ magnitudes above the WHDF magnitude limit. To
determine the galaxy counts below the limit of the WHDF survey we
assume that the counts in the bins $2-3$ magnitudes below the survey
limit follow the same linear trend (in the log space) as the counts in
the brighter bins used in the fit. We finally take into account the
incompleteness of our own data and correct the derived galaxy number
counts using the completeness curves presented in
Figure~\ref{fig3}. The estimated mean number of unresolved background
galaxies over all fields is $36\pm7$ per arcmin$^{-2}$ for our RGB
selection box and $15\pm8$ per arcmin$^{-2}$ for the region used to
select the MS stars. The errors reflect both the variation of the
estimated galaxy counts between the three fields and the cosmic
variance.  The latter was estimated using \textsc{GalaxyCount}, which
for given galaxy number counts in a given magnitude range calculates
the expected variance. The star counts profiles before and after
subtraction of the background galaxy population are shown in
Figure~\ref{fig9}, in top and bottom panels, respectively.  Error bars
in the bottom panel are a combination of errors quoted above and
Poisson uncertainties.  Under the assumption that the slope of the
WHDF number counts changes at the survey magnitude limit to a value
twice that derived from the bins $3$ magnitudes below the survey
limit, we derive the galaxy counts as $80\pm15$ and $29\pm14$ per
arcmin$^{-2}$ for our RGB and MS selection, respectively.  After the
subtraction of the newly determined galaxy counts, the old stellar
disk remains purely exponential but falls below the detection limit at
$\sim20$ arcmin.  However, these higher counts are very likely to be
an unrealistically high estimate for the contribution from the
background galaxy population, given that they are significantly higher
than our MS star counts in the outermost two thirds of our MS radial
profile.

As the Figure~\ref{fig9} clearly shows, while the old stellar disk
extends out to $25'$, corresponding to $15$ kpc, the young main
sequence stellar population is more centrally concentrated and is only
detected out to $17'$ ($10$ kpc) along the southern end of the major
axis. The disk, as traced by the old stellar population, does not
exhibit a break in surface brightness profile once the star counts
have been corrected for the contamination by unresolved background
galaxies. This complements our finding of an unbroken exponential disk
in late type galaxy NGC~300 out to the distances corresponding to
about $10$ disk scale lengths \citep{blandhawthorn05}.

\subsection{Surface Brightness Profile}

\begin{figure}[]
\plotone{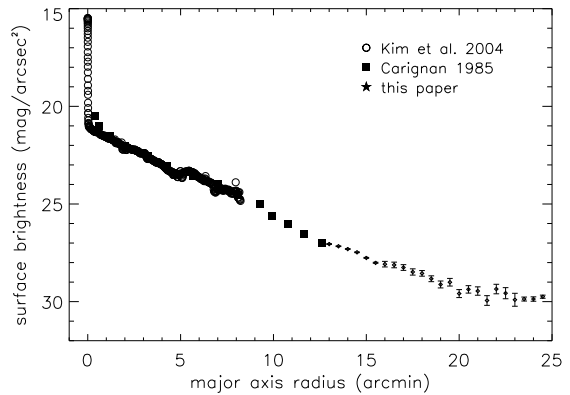}
\caption{Surface brightness profile of NGC~300. Open circles are
  I-band surface photometry data from \citet{kim04}; the data points
  have been shifted downward by $1.2$ mag. Full squares are B$_J$
  surface photometry measurements taken from \citet{carignan85} and
  have been shifted upward by 0.4 mag. The star symbols are our
  surface brightness measurements derived from star
  counts.\label{fig10}}
\end{figure}

In Figure~\ref{fig10} we derive the surface brightness profile of the
outer disk of NGC~300 from the $i'$-band starcounts of the old RGB
stars.  The inner data points in this figure are taken from the
surface photometry studies by \citet{kim04} and \citet{carignan85}.
The outer disk surface brightness is calculated in $0.5'$ annuli by
summing the flux of all stars in a given annulus and correcting for
the radial completeness, number of pixels in the annulus and the pixel
scale to obtain surface brightness in the units of mag arcsec$^{-2}$.
We correct for the background galaxy contamination by randomly
removing from each radial bin a number of objects corresponding to our
estimate of the galaxy number counts (36 per arcmin$^{-2}$) corrected
for the area of the annuli.  We repeat this process by drawing 100
random ``background galaxy-corrected'' samples from our full stellar
catalogue and use this to calculate the final value of surface
brightness in each radial bin, and estimate its error bars (as
standard deviations of the full set of 100 samples).  Finally, we
correct for the light missed due to the fact that we detect only the
brightest stars in the outskirts of NGC~300 and the inclination of the
galaxy.  In \citet{blandhawthorn05} we estimate the fraction of
missing light to be $\sim45\%$ for a study of a similar photometric
depth.  This translates to intrinsic surface brightness 0.65 mag
brighter than what we derive here.  The galaxy inclination
($i=42^\circ$) somewhat counteracts this effect; if the galaxy was
viewed face-on the resulting surface brightness would be 0.32 mag
fainter.  Therefore, to account for both effects the calculated
surface brightness has to be shifted upward by 0.33 mag.

The surface brightness profile in Figure~\ref{fig10} emphasizes the
power of resolved star counts over the traditional diffuse light
imaging for reaching down to very low surface brightness levels
necessary for studying the extremely faint outskirts of galaxy disks.

\subsection{Metallicity Distribution Function}
\label{sec:mdf}

Given the color-magnitude diagram and the distance to NGC~300, we are
able to compute the metallicity distribution function of the stellar
populations in its outskirts.  Theoretical isochrones show that
metallicity (rather than age) is the primary factor affecting the
color of the red giant branch (\citealp{vandenberg06}; see also
Figure 5 and 6 of \citealp{harris99}); for example, at absolute
magnitude $M_I=-3$ the $0.2$ mag blueward shift in $V-I$ can be
achieved by a decrease in age from $13$ to $6$ Gyr, or by a relatively
small decrease in [Fe/H] from $-1.0$ to $-1.2$
\citep{mouhcine05}. This high sensitivity of photometric properties on
metallicity allows us to derive the metallicity distribution function
from the photometry of the red giant branch stars.

\begin{figure}[]
\plotone{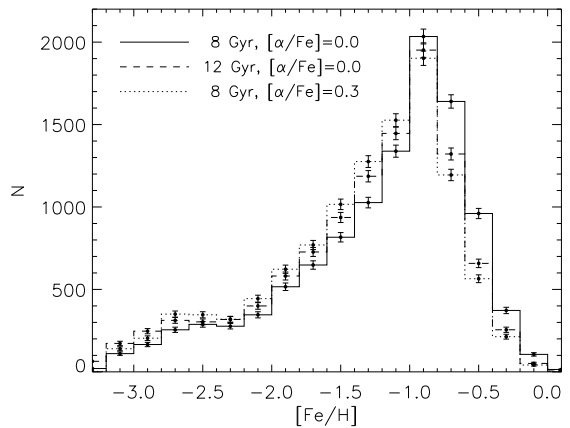}
\caption{Metallicity distribution function of NGC~300 stars calculated
  using the Victoria-Regina stellar evolutionary tracks with the age
  of 8 Gyr and no $\alpha$-enhancement. Also shown are the metallicity
  distribution functions calculated assuming the stellar ages of $12$
  Gyr and [$\alpha$/Fe]$=0.0$ (dashed line) and an $8$ Gyr old stellar
  population with [$\alpha$/Fe]$=0.3 $ (dotted line).  If the stellar
  population in the outskirts of NGC~300 is $\alpha$-enhanced or older
  than the assumed age of $8$ Gyr, our procedure will have slightly
  overestimated the number of stars at the high-metallicity end, but
  without a significant effect on the mean metallicity of the
  population. \label{fig11}}
\end{figure}

To derive the metallicity distribution function we convert the
observed distance- and reddening-corrected stellar photometry to
metallicity on a star-by-star basis. We superimpose on observed
color-magnitude diagram the stellar evolutionary tracks of
\citet{vandenberg06} and interpolate between them to derive an
estimate of star's metallicity (Figure~\ref{fig8}). The model grid
consist of 16 finely spaced red giant tracks for a 8 Gyr stellar
population without $\alpha$-enhancement, covering the range of
metallicities from [Fe/H]$=-2.31$ to $0.00$ in the steps of
approximately $0.1$ dex.  We define a region in ($g'-i'$,$i'$) which
we expect to be populated primarily by red giant branch stars 
and only consider the stars within this selection box when
calculating metallicities (Figure~\ref{fig8}).

In computing the metallicity distribution function we choose to use
the theoretical stellar evolutionary tracks rather than empirical globular
cluster fiducials. The grid of red giant branch tracks from
\citet{vandenberg06} on which we base our metallicity estimates is much
more finely spaced than often used Milky Way globular cluster
fiducials by \citet{saviane00}; as shown by \citet{harris99} and
\citet{harrisharris00} the coarseness of the model grid may introduce
artifacts in the calculated metallicity distribution function. By  
inspection of the Figure~\ref{fig8} it is clear that the
stars in the outer regions of NGC~300 are metal poor and there is no
need for adding stellar tracks more metal rich than those present in
\citet{vandenberg06} grid. 

In order to take into account the incompleteness effects we compute
the completeness-corrected metallicity distribution function by
counting each star $(C_{g'}C_{i'})^{-1}$ times, where $C_{g'}$ and
$C_{i'}$ are completeness in $g'$ and $i'$ at a given point in the
CMD.  We only consider stars which fall above $50\%$ completeness
limit in both $g'$ and $i'$.  For fainter stars the completeness
correction factor becomes too large and not necessarily a realistic
representation of the incompleteness effects which are poorly
understood at these faint magnitudes.

The presence of stars of different ages will affect our metallicity
estimates which utilize the $8$ Gyr stellar tracks; the effect is
illustrated in Figure~\ref{fig11}. The solid line is the metallicity
distribution function calculated using the stellar tracks for the $8$
Gyr stars with no $\alpha$-enhancement.  Also shown are the
metallicity distributions constructed assuming an older age ($12$ Gyr,
dashed line) and different value for $\alpha$-enhancement
([$\alpha$/Fe]$=0.3 $, dotted line). If the true stellar population is
older or more $\alpha$-enhanced than our adopted stellar evolutionary
tracks, this procedure will have overestimated the number of
high-metallicity stars, and similarly, underestimated the contribution
of low-metallicity stars, but the effect on the peak metallicity of
the stellar population will not be significant.

\subsubsection{Metallicity Gradient}

\begin{figure}[]
\plotone{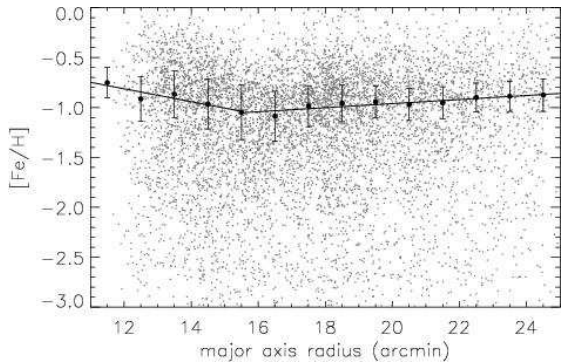}
\caption{Metallicity as a function of radius for all stars in the
  outer disk of NGC~300. Full circles are the median metallicities in
  $1'$ bins.  Solid lines represent the metallicity gradient and are a
  linear fit to the median metallicities in the range $R<15.5'$ and
  $R>15.5'$.  Error bars represent the mean uncertainty in [Fe/H] in
  each bin.}
\label{fig12}   
\end{figure}

In Figure~\ref{fig12} we plot metallicity as a function of deprojected
galactocentric radius for all stars in the outer disk of NGC~300.  
Preliminary analysis of stellar metallicities has shown that
metallicity decreases with radius out to about $15.5'$ after which the
gradient becomes mildly positive.  We therefore divide the data into
$R<15.5'$ and $R>15.5'$ ($9.3$ kpc) bins and independently calculate
the metallicity gradient (in dex kpc$^{-1}$) for each region, by
fitting linear functions to median metallicities in $1'$ wide radial
bins:

\[ R<15.5' : \mathrm{[Fe/H]} = -0.11(\pm0.02) R -0.04(\pm0.15)\]
\[ R>15.5' : \mathrm{[Fe/H]} = 0.034(\pm0.004) R -1.37(\pm0.06)\]

Despite the large scatter in calculated metallicities, we clearly
detect both the negative abundance gradient out to $\sim9.3$ kpc and a
change in the trend outwards of $9.3$ kpc.  

\begin{figure}[]
\plotone{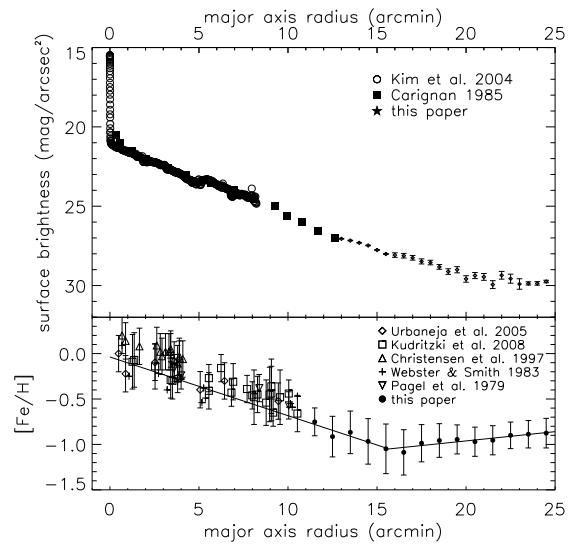}
\caption{{\it Top:} Surface brightness profile of NGC~300. {\it
    Bottom:} Metallicity gradient in the disk of NGC~300. Inner disk
    data points are from the spectroscopic abundance studies listed in
    the figure.  Filled circles are outer disk binned mean
    metallicities and solid lines are the linear fits to these values
    as calculated in Figure~\ref{fig12}.
\label{fig13}}
\end{figure}

Figure~\ref{fig13} shows the pure exponential surface brightness
profile of NGC~300 and its metallicity gradient on the same radial
scale.  In the bottom panel of this figure we compile the available
data on the inner disk metallicities from the literature.  All
metallicities except the recent \citet{kudritzki08} results refer to
oxygen metallicities only, and all except \citet{kudritzki08} and
\citet{urbaneja05} metallicities are derived using the \HII-region
emission lines.

The reasons for the metallicities we derive being seemingly too low
compared to the spectroscopic metallicities in the central regions of
NGC~300 are two-fold.  Firstly, the inner disk metallicities in the
Figure~\ref{fig13} either represent gas, rather than stellar,
metallicities, or were measured from young blue supergiants, both of
which are expected to have higher metallicities than the older red
giant branch stars at the same radii.  Secondly, as can be seen from
Figure~\ref{fig12}, the bulk of the stars populate the
high-metallicity end of the plot and it is the low-metallicity tail of
the abundance distribution function (see Figure~\ref{fig11}) that is
responsible for lower mean metallicities. Since we base our discussion
primarily on the overall shape of the metallicity gradient, rather
than the absolute value of metallicities this effect will not
influence our conclusions.

In order to test the robustness of the derived abundance gradient, we
performed a series of tests in which we randomly removed from the
sample (i) a constant number of stars, corrected for the different
areas of elliptic annuli, or (ii) a number of stars proportional to
the number density of stars in a given radial annuli.  These
procedures mimic the effect of correcting for the (i) constant
contamination (e.g. background galaxies, stellar halo) and (ii)
radially dependent contamination (e.g. thick disk).  Only after
artificially removing ~50\% of stars in each bin are we able to derive
a flat abundance gradient over the full radial extent sampled by our
data.  However, there is no realistic mechanism that could account for
a 50\% contamination level within our RGB selection region and hence
we consider our derived abundance gradient to be robust.

\section{DISCUSSION}
\label{sec:discussion}

In this paper we revisit NGC~300 to investigate the faint stellar
population in its outer disk.  We confirm the earlier finding
\citep{blandhawthorn05} of the extended stellar disk in NGC~300 with
the surface brightness declining exponentially out to $\sim25'$, or
about $10$ disk scale lengths (\S\ref{sec:radialprofile}).  Only three
other disk so far have been traced down in resolved stars to the same
effective surface brightness levels -- that of the Galaxy
\citep{ibata03}, M31 \citep{irwin05} and M33
\citep{ferguson07}. Except for the case of a ``sub-exponential'' disk
of M33 the other three disks are either pure exponentials or
``super-exponential'' and are detected down to the effective surface
brightness levels below $\sim30$ mag arcsec$^{-2}$. The existence of
stars at these extremely low surface densities in the outer galactic
disks is puzzling. We want to learn what the nature of the faint
stellar population in NGC~300 is and whether its outer disk is young
or old. The answers to these questions will have important
implications for the scenarios for the formation of spiral
galaxies. With the multi-band imaging in hand, we are now able to
assess the ages and metallicities of stars in the outskirts of this
late type spiral.

From the ($g'-i'$,$i'$) color magnitude diagrams of NGC~300 presented
in \S\ref{sec:cmd} and the star counts profiles of RGB and MS stars
(Figure~\ref{fig9}) we find that its outer disk is
predominantly old. Old outer disks are in disagreement with the
inside-out picture of spiral galaxy formation in which the inner disk
is built up on shorter timescales than the outer disk and young stars
are expected to inhabit the larger radii
\citep{larson76, matteuccifrancois89,chiappini97,naabostriker06}.
This is however not the first time the old stellar population has been
discovered at large galactocentric distances. 
\citet{fergusonjohnson01} use WFPC2 on board HST to observe a field on
the major axis of M31 at the radius corresponding to 3 disk scale
lengths.  Using the mean I-band magnitude of the red clump as an age
indicator they find that the mean age of the stellar population in the
outskirts of M31 is $>8$ Gyr.  In their studies of the outermost
regions of M33, \citet{galleti04} and \citet{barker07} both find that
the young main sequence stars appear to be more centrally concentrated
and hence have smaller scale lengths than the older red giant branch
population. The disks that have formed inside-out however have stellar
populations which get progressively younger with radius and the oldest
disk stars have the smallest scale lengths, in contrast to what is
observed in M33
\citep{matteuccifrancois89,chiappini97,galleti04,barker07}. 

It may be possible to reconcile the two results in the context of
models by \citet{roskar08}.  In their simulations of spiral galaxy
formation the old outer disks arise as a consequence of stellar
migrations from the inner disk.  Star formation rate density falls off
sharply at the radius of a break in the radial profile, but the stars
are found in the outer disk at least $\sim5$ kpc beyond the break.
They find that about $85\%$ of these stars have formed in the inner
disk and subsequently migrated outwards, beyond the truncation
radius. The break in the surface density coincides with the minimum
mean stellar age in the disk.  Inwards of the break younger stars
populate the larger radii, but the trend is reversed once we cross the
point of truncation. In this scenario, the outermost regions of disks
are old because the old stars have had the most time to migrate
farthest out in the disk. While the positive age gradient in
\citet{roskar08} models could succeed in explaining the
\citet{galleti04} and \citet{barker07} results for the outer disk of
M33, it is not clear whether the same mechanism would produce the old
outer disk in the purely exponential disk of NGC~300.

The question that arises here is whether what we observe as an
extended stellar disk is in fact a halo component in NGC~300. 
\citet{tikhonov05} use HST/ACS to study the outer disk of NGC~300 and
attribute the red giants found in their outermost field at $13'$ to a
halo population.  However, our surface brightness profile of NGC~300 in
Figure~\ref{fig10} confirms that the outer disk
indeed follows the same exponential profile as the inner disk and
hence is not a distinct population. The possible reasons for a
discrepant result reported by \citet{tikhonov05} are two-fold.  Their
outer field spans only $2'$ with large variations of star counts
within the field, and they do not account for the contribution to the
star counts by faint background galaxies.  Correcting for the
contamination from the unresolved galaxy population would have the
effect of steepening the profile in the outer field and making its
slope more similar to the slope derived from their innermost fields.

Using the color of the stars on the red giant branch as a proxy for
their metallicity, we derive the metallicity distribution function of
the stars in the outskirts of NGC~300 and find that metallicity
gradient changes slope at a radius of $15.5'$ ($\sim9.3$ kpc).  Does
this mark the end of the disk in NGC~300 and a transition to an old
halo?  The argument against halo domination at large galactocentric
radii is the relatively high metallicity in the outermost
$10'$. Typical stellar halos are more metal-poor than what we find for
the outer disk of NGC~300. In the Local Group, the halos of the Galaxy
\citep{beers05,carollo07,ivezic08} and M31 \citep{chapman06,koch07}
have metallicities of the order of $\mathrm{[Fe/H]}=-1.5$, or lower.
Even the less massive, almost bulge-less M33, believed to be a local
analogue of NGC~300, has a metal-poor halo with $\mathrm{[Fe/H]}=-1.3$
to $-1.5$ \citep{brooks04,mcconnachie07}, and a more metal-rich disk
with the photometric metallicity of [Fe/H]$\approx-0.9$
\citep{mcconnachie07}, similar to the outer disk metallicities we
derive. We also note here that stellar halos are usually found to
exhibit a negative metallicity gradient
\citep{harris96,parmentier00,koch07}, which is in disagreement with
our positive or flat gradient in the outer disk.  However, due to the
limited radial extent of our data, this is not as strong of an
argument against the halo domination at large radii as are the high
metallicities found in our outermost field.

Negative abundance gradients are a common feature of disk galaxies.
The dependence of yield, star formation and gas infall on the
galactocentric radius in chemical evolution models reproduces the
gradients \citep{goetzkoeppen92,matteuccifrancois89}, which are
shallow and with a large but uniform dispersion.  A metallicity trend
in which the decline in abundances with radius is followed by a
metallicity plateau, and which we find in the outer disk of NGC~300,
may be a general property of spiral disks.  It has so far been
observed in all of the galaxies for which the deep stellar photometry
is available -- M31, M33 and the Galaxy.  \citet{worthey05} use
archival HST images of a number of fields spanning the galactocentric
distances out to $50$ kpc to derive abundance gradient in the disk of
M31.  While the inner disk exhibits a negative metallicity gradient,
as expected from the chemical evolution models, the gradient
disappears at $\sim25$ kpc with the metallicity remaining nearly
constant in the outer disk at
$\mathrm{[M/H]}\approx-0.5$. \citet{yong05}, \citet{carney05} and
\citet{yong06} find a similar metallicity plateau at the radius of
$10-12$ kpc in the outer disk of the Galaxy. They also observe the
$\alpha$-enhanced outer disk with
$\mathrm{[\alpha/Fe]}\approx0.2$. Similar results are those of
\citet{twarog97} who find the break in the metallicity gradient of
open clusters at R$\approx10$ kpc, with the constant metallicity at
larger galactocentric distances. In their deep ACS study of the M33
outer disk, \citet{barker07} see an indication for the flattening in
the metallicity gradient beyond the radius of $50'$.  The same effect
has also been recently detected through the oxygen abundances of \HII
regions in M83 \citep{bresolin09}.

\begin{figure}[]
\plotone{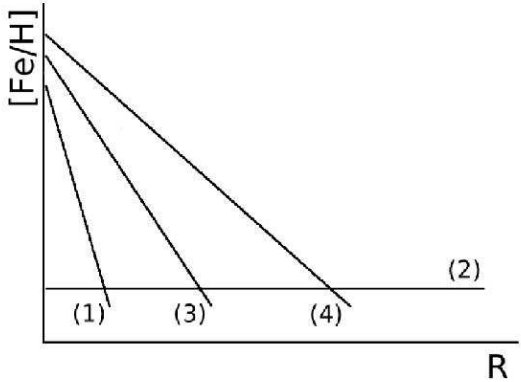}
\caption{The illustration of a working model for the evolution of
  abundance gradients in disks.  (1) Central spheroid is formed at
  high redshift and intense star formation establishes the steep
  metallicity gradient. (2) The metal-poor and uniform-metallicity
  disk is assembled at $z\sim2$. (3) Star formation migrates to
  progressively larger radii and enriches the initially
  low-metallicity regions. (4) As a consequence, the abundance
  gradient continues to flatten. \label{fig14}}
\end{figure}

We present here two scenarios to explain the metallicity plateau (or
upturn) in the outer disks of spirals. 

In the radial mixing scenario, the mechanism described by
\citet{sellwoodbinney02} is responsible for transporting the stars
within the disk while preserving the nearly circular orbits and
exponential light profile. In \citet{blandhawthorn05} we show that the
outer disk of NGC~300 is a high-Q environment (with the Toomre Q
parameter of $\sim5\pm2$) and as such is expected not to be responsive
to gravitational disturbances, such as star formation or spiral waves.
If however there is a mechanism that could lower the estimated value
of Q (e.g. presence of cold gas that is unaccounted for in our earlier
calculation of Q which estimates the surface density from the \HI
observations), scattering of stars by spiral waves
\citep{sellwoodpreto02} could possibly explain the existence of stars
in these diffuse regions. Potentially, stellar migrations could also
explain the flattening or upturn we observe in the metallicity
gradient of NGC~300.  \citet{roskar08} reproduce a similar change in
trend of the stellar age with radius. However, scattering over such
large distances ($5-8$ kpc) would require spiral waves much stronger
than those considered by \citet{sellwoodbinney02}.

The accretion scenario is illustrated in Figure~\ref{fig14}.  In this
picture, the central spheroid is assembled first in the early Universe
($z>4$).  The first generations of stars that form establish a steep
abundance gradient in the spheroid (1), reaching the metallicities of
around solar. This is supported by solar metallicities seen in QSOs at
high redshifts \citep[and references therein]{hamannferland99} and
steep abundance gradient in the old stellar bulge of the Galaxy
\citetext{\citealp{zoccali08,feltzinggilmore00,frogel99,minniti95},
  but see \citealp{ramirez00,rich07}}.  The disk component is
established at $z\sim2$ \citep{wolfe05}.  The initial disk is
metal-poor and with uniform abundance (2), reflecting the
metallicities in the IGM at the formation redshift.  As the galaxy is
built in the inside-out fashion, the extent of star formation migrates
towards the outer disk and the parts of the disk that previously had
pristine abundances are now forming metals (3).  This is a consequence
of the gas and stellar density building up in the outer disk. As a
result, regions of the disk at progressively larger radii become
enriched and abundance gradient flattens with time (4).  Observations
of a variety of chemical gradients in PNe \citep{maciel03,maciel06},
open clusters \citep{friel02,chen03}, cepheids, OB associations
\citep{maciel05,dafloncunha04} and \HII regions \citep{maciel07} have
shown that older objects exhibit steeper abundance gradients resulting
in chemical gradients that become flatter with time.
We note however that while some of the chemical evolution models that
incorporate inside-out growth of galactic disks are able to reproduce
this behavior
\citep{molla97,boissierprantzos99,portinarichiosi99,hou00}, others
predict steepening of gradient with time \citep{tosi88,chiappini01}.
This demonstrates the sensitivity of abundance gradients on adopted
form of the radial dependence for the star formation and gas infall.

A simple consequence of this picture is that there exists a radius in a
galactic disk at which a transition occurs between the star forming
disk with a negative abundance gradient and a uniform-abundance initial
disk, as observed here in the case of NGC~300, as well as in the disks
of the Galaxy \citep{yong05,carney05,yong06} and M31
\citep{worthey05}.  The outer disk abundances reflect the local
metallicity floor, which is supported by enhanced [$\alpha$/Fe] ratios
and low [Fe/H] both in external gas locally
\citep{collins03,collins07} and in the intergalactic medium at high
redshift \citep{prochaska03apjl,dessauges04,dessauges06}.

The accretion scenario suggests that stars today were born roughly in
situ, whereas the radial mixing scenario implies that stars largely
migrate away from where they were formed. Both of these have
consequences for future studies of disk formation.

\acknowledgments

We thank the anonymous referee for helpful comments, James Binney for
useful discussion and Peter Stetson for providing his DAOPHOT package.

{\it Facilities:} \facility{Gemini (GMOS)}.

\newpage

\end{document}